\documentclass[aps,prb,twocolumn,groupedaddress,showpacs,floatfix]{revtex4-1}
\pdfoutput=1

\usepackage{graphicx}
\usepackage{dcolumn}
\usepackage{bm}
\usepackage{amsmath}
\DeclareMathOperator{\Tr}{Tr}

\begin{document}
\title{Beyond the constant-mass Dirac physics: Solitons, charge fractionization, and 
the emergence of topological insulators in graphene rings}

\author{Constantine Yannouleas}
\email{Constantine.Yannouleas@physics.gatech.edu}
\author{Igor Romanovsky}
\email{Igor.Romanovsky@physics.gatech.edu}
\author{Uzi Landman}
\email{Uzi.Landman@physics.gatech.edu}

\affiliation{School of Physics, Georgia Institute of Technology,
             Atlanta, Georgia 30332-0430}

\date{8 July 2013; Physical Review B, {\bf in press}}

\begin{abstract}
The doubly-connected polygonal geometry of planar graphene rings is found to bring forth topological
configurations for accessing nontrivial relativistic quantum field (RQF) theory models that carry 
beyond the constant-mass Dirac-fermion theory.  These include generation of sign-alternating masses, 
solitonic excitations, and charge fractionization. The work integrates a RQF Lagrangian formulation 
with numerical tight-binding Aharonov-Bohm electronic spectra and the generalized 
position-dependent-mass Dirac equation. In contrast to armchair graphene rings (aGRGs) with pure
metallic arms, certain classes of aGRGs with semiconducting arms, as well as with mixed 
metallic-semiconducting ones, are shown to exhibit properties of one-dimensional nontrivial 
topological insulators. This further reveals an alternative direction for realizing a graphene-based
nontrivial topological insulator through the manipulation of the honeycomb lattice geometry, without
a spin-orbit contribution.
\end{abstract}

\pacs{73.22.-f, 03.70.+k, 05.45.Yv}

\maketitle

\section{Introduction}

Research endeavors aiming at realization 
\cite{geim04,kats06,geim09,haka10,zhan11,zhan12,tarr12,mano12,been13}
of vaunted relativistic quantum field (RQF) behavior 
\cite{griffbook} in ``low-energy'' laboratory setups were spawned by the isolation of graphene, 
\cite{geim04,geim09} whose low-energy excitations behave as massless Dirac-Weyl (DW) fermions 
(moving with a Fermi velocity $v_F$ instead of the speed of light $c$; $v_F \sim c/300$), offering 
a link to quantum electrodynamics \cite{geim09,kats06,kats07,chou13} (e.g., Klein tunneling and 
Zitterbewegung). 

Here we show that planar polygonal graphene rings with armchair edge terminations
(aGRGs) can provide an as-yet unexplored condensed-matter bridge to high-energy particle physics 
beyond both the massless Dirac-Weyl and the constant-mass Dirac fermions. Due to their doubly 
connected topology [supporting Aharonov-Bohm (AB) physics \cite{imry83}] aGRGs bring forth 
condensed-matter realizations for accessing acclaimed one-dimensional (1D) RQF models involving 
the emergence of position-dependent masses and consideration of interconnected vacua (or 
topological domains). As a function of the ring's arm width, one finds two general outcomes: 
(I) Formation of soliton/anti-soliton 
fermionic complexes \cite{jasc81,heeg88,jakl83} studied in the context of charge fractionization 
\cite{jare76} and the physics of {\it trans\/}-polyacetylene, \cite{jasc81,heeg88} and (II) Formation 
of fermion bags introduced in the context of the nuclear hadronic $\sigma$ model \cite{camp76} and in 
investigations of non-trivial Higgs-field mass acquisition for heavy quarks. \cite{mapa90}

A principal result of our study is that it reveals an emergent alternative direction for realizing 
a graphene-based nontrivial topological insulator \cite{haka10,zhan11} (TI) through the manipulation 
of the honeycomb lattice geometry, without a spin-orbit contribution. In particular, in contrast to 
armchair graphene rings with pure metallic arms, certain classes of aGRGs with semiconducting
arms, as well as with mixed metallic-semiconducting ones, are shown to exhibit properties of 
one-dimensional nontrivial TIs. 

\section{Methodology}

The energy of a particle (with onedimensional momentum $p_x$) is given by the Einstein
relativistic relation $E=\sqrt{ (p_x v_F)^2+({\cal M} v_F^2)^2 }$, where ${\cal M}$ is the
rest mass. In gapped graphene or graphene systems, the mass parameter is related to the
particle-hole energy gap, $\Delta$, as ${\cal M}=\Delta/(2 v_F^2)$. In RQF theory, the mass of 
elementary particles is imparted through interaction with a scalar field known as the Higgs field.
Accordingly, the mass ${\cal M}$ is replaced by a position-dependent Higgs field 
$\phi(x) \equiv m(x)$, to which the relativistic fermionic field $\Psi(x)$ couples through the Yukawa 
Lagrangian \cite{mapa90,roma13} ${\cal L}_Y = -\phi \Psi^\dagger \beta \Psi$ ($\beta$ being a Pauli 
matrix). In the elementary-particles Standard Model, \cite{griffbook} such coupling 
is responsible for the masses of quarks and leptons. For $\phi(x) \equiv \phi_0$ (constant)
${\cal M} v_F^2=\phi_0$, and the massive fermion Dirac theory is recovered.

We exploit the generalized Dirac physics governed by a total Lagrangian density ${\cal L}=
{\cal L}_f + {\cal L}_\phi$, where the fermionic part is given by 
\begin{equation}
{\cal L}_f = - i \hbar \Psi^\dagger \frac{\partial}{\partial t} \Psi
- i \hbar v_F \Psi^\dagger \alpha \frac{\partial}{\partial x} \Psi +{\cal L}_Y ,
\label{lagrf}
\end{equation}
and the scalar-field part has the form
\begin{equation}
{\cal L}_\phi =  - \frac{1}{2} (\frac{\partial \phi}{\partial x})^2 - 
\frac{\xi}{4} (\phi^2 - \phi_0^2)^2,
\label{lagrphi}
\end{equation}
with the potential $V(\phi)$ (second term) assumed to have a double-well $\phi^4$ form;
$\xi$ and $\phi_0$ are constants.

Henceforth, the Dirac equation is generalized as
\begin{equation}
E \Psi + i \hbar v_F \alpha \frac{\partial \Psi}{\partial x} - \beta \phi(x) \Psi=0.
\label{direq}
\end{equation} 
In one dimension, the fermion field is a two-component spinor $\Psi = (\psi_u, \psi_l)^T$;
$u$ and $l$ stand, respectively, for the upper and lower component and
$\alpha$ and $\beta$ can be any two of the three Pauli matrices. 

Each arm of a polygonal ring can be viewed as an approximation of an armchair graphene nanoribbon
(aGR). The excitations of an infinite aGR are described by the 1D massive Dirac 
equation, see Eq.\ (\ref{direq}) with $\alpha=\sigma_2$, $\beta=\sigma_1$, and 
$\phi(x) \equiv \phi_0 = \Delta/2 \equiv |t_1-t_2|$. The two (in general) unequal hopping parameters 
$t_1$ and $t_2$ are associated with an effective 1D tight-binding problem (see Appendix \ref{tbm}) 
and are given \cite{zhen07} by 
$t_1=-2 t \cos[p \pi/({\cal N}_W+1)]$, $p=1,2,\ldots,{\cal N}_W$ and $t_2=-t$; ${\cal N}_W$ is 
the number of carbon atoms specifying the width of the nanoribbon and $t=2.7$ eV is the hopping 
parameter for 2D graphene. The effective \cite{zhen07} TB Hamiltonian of an aGR has a 
form similar to that used in {\it trans}-polyacetylene (a single chain of carbon atoms). In 
{\it trans}-polyacetylene, the inequality of $t_1$ and $t_2$ (referred to as dimerization) is a 
consequence of a Peierls distortion induced by the electron-phonon coupling. For
an aGR, this inequality is a topological effect associated with 
the geometry of the edge and the width of the ribbon. We recall that as a function of their width, 
${\cal N}_W$, the armchair graphene nanoribbons fall into three classes: (I) ${\cal N}_W=3l$ 
(semiconducting, $\Delta>0$), (II) ${\cal N}_W=3l+1$ (semiconducting, $\Delta>0$), and (III) 
${\cal N}_W=3l+2$ (metallic $\Delta=0$), $l=1,2,3,\ldots$.  

We adapt the ``crystal'' approach \cite{imry83} to the AB effect, and introduce a virtual
Dirac-Kronig-Penney \cite{mcke87} (DKP) relativistic superlattice (see Appendix \ref{dkpsm}). Charged 
fermions in a perpendicular magnetic field circulating around the ring behave like electrons in a 
spatially periodic structure (period ${\cal D}$) with the magnetic flux $\Phi/\Phi_0$ ($\Phi_0=hc/e$) 
playing the role of the Bloch wave vector $k$, i.e., $2 \pi \Phi/\Phi_0=k {\cal D}$ [see the cosine 
term in Eq.\ (\ref{disrel})]. 

\begin{figure}
\centering\includegraphics[width=7.0cm]{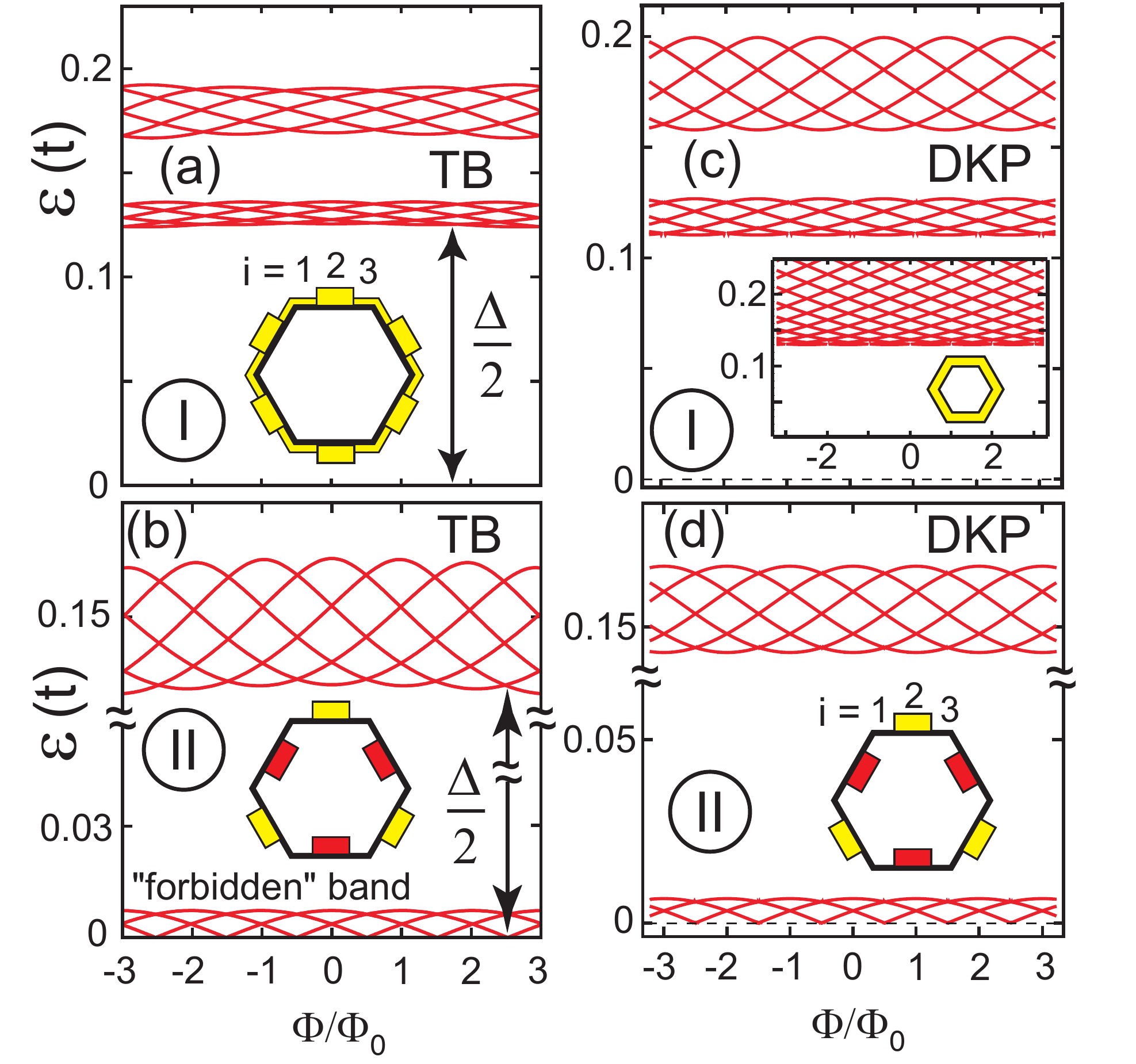}
\caption{
{\it Aharonov-Bohm spectra for hexagonal armchair graphene rings.\/}
(a) Tight-binding spectrum for a class-I nanoring with ${\cal N}_W=15$. (b) TB spectrum for a 
class-II nanoring with ${\cal N}_W=16$. The armchair graphene rings are semiconducting (${\cal N}_W=15$)
and metallic at $\Phi=(\pm j+1/2) \Phi_0$, $j=1,2,3,\ldots$ (${\cal N}_W=16$).  
The two lowest-in-energy six-membered bands are shown. The hole states (with 
$\varepsilon < 0$, not shown) are symmetric to the particle states (with $\varepsilon > 0$). 
Insets: schematics of the Higgs fields $\phi(x)$ employed in the DKP modeling. $\phi(x)$ is 
approximated by steplike functions $m_i^{(n)}$; $i$ counts the three regions of each arm
($L_1^{(n)}=L_3^{(n)}=a$ and $L_2^{(n)}=b$), and
$n$ ($n=1,\ldots,6$) counts the hexagon's arms. The non-zero (constant) variable-mass values of
$\phi(x)$ are indicated by yellow (red) color when positive (negative). 
These resulting DKP spectra [(c) and (d)] reproduce the TB 
ones in (a) and (b), respectively. 
The parameters used in the DKP modeling are: (c) $a=2 a_0$, $b=28 a_0$, $m_1^{(n)}=m_3^{(n)}=0.06
t/v_F^2$, $m_2^{(n)}=0.13 t/v_F^2$ [see corresponding schematic inset in (a)] 
and (d)  $a=7 a_0$, $b=15 a_0$, $m_1^{(n)}=m_3^{(n)}=0$, $m_2^{(n)}=(-1)^n m_0$ with 
$m_0=0.18 t/v_F^2$ [see schematic inset]. 
The inset in (c) shows the spectrum for a free massive Dirac fermion with a constant
mass ${\cal M}=0.13 t/v_F^2$.
Note the six-membered braided bands and the ``forbidden'' band [within the gap, in (b) and (d)].
$a_0=0.246$ nm is the graphene lattice constant and $t=2.7$ eV is the hopping parameter.
}
\label{fig1}
\end{figure}  

\section{Results}
\subsection{Rings with semiconducting arms}

Naturally, nanorings with arms made of nanoribbon segments belonging to the semiconducting classes I 
and II may be expected to exhibit a particle-hole gap (particle-antiparticle gap in RQF theory). 
Indeed this is found for class I aGRGs [see gap $\Delta$ in Fig.\ \ref{fig1}(a)]. Suprisingly, the 
class II nanorings demonstrate a different behavior, showing a ``forbidden'' band in the middle of the 
gap region [see Fig.\ \ref{fig1}(b)]. This forbidden band is dissected by the zero-energy axis and its 
members cross this axis at regular magnetic-flux intervals 
$\Phi=(\pm j+1/2) \Phi_0$, $j=1,2,3, \ldots$, manifesting semi-metallic behavior. 

This behavior of class II aGRGs can be explained through analogies with RQF theoretical models, 
describing single zero-energy fermionic solitons with fractional charge \cite{jare76,jasc81} or
their modifications when forming soliton/anti-soliton systems. \cite{jakl83,jasc81} 
(A solution of the equation of motion corresponding to Eq.\ (\ref{lagrphi}), is a $Z_2$
kink soliton, $\phi_k(x)$. The solution of Eq.\ (\ref{direq}) with $\phi=\phi_k(x)$ is the 
fermionic soliton.) We model the hexagonal ring with the use of a continuous 1D Kronig-Penney 
\cite{mcke87,note1} model (see Appendix \ref{dkpsm}) based on the generalized Dirac equation 
(\ref{direq}), allowing variation of the scalar 
field $\phi(x)$ along the ring's arms. We find that the DKP model reproduces [see Fig.\ 
\ref{fig1}(d)] the spectrum of the class-II ring (including the forbidden band) 
when considering alternating masses $\pm m_0$ associated with contiguous arms 
[see inset in Fig.\ \ref{fig1}(b)].

The sign-alternating mass regions, separated by six regions of vanishing mass centered at the corners,
corresponds to a Higgs field composed of a train (or a so-called crystal \cite{thie06,dunn08,taka13})
of three kink/antikink soliton pairs. In analogy with the physics of
{\it trans\/}-polyacetylene, the positive and negative masses correspond to two 
degenerate domains associated with the two possible dimerization patterns \cite{jasc81,heeg88} 
$\ldots -t_1-t_2-t_1-t_2- \ldots$ and $\ldots -t_2-t_1-t_2-t_1- \ldots$, which are possible in a 
single-atom chain. The transition zones between the two domains (here the corners of the hexagonal 
ring) are referred to as the domain walls.

For a single soliton, a (precise) zero-energy fermionic excitation emerges, 
localized at the domain wall. In the case of soliton-antisoliton pairs, paired energy levels with
small positive and negative values appear within the gap. The TB spectrum 
in Fig.\ \ref{fig1}(b) exhibits a forbidden band of six paired $+/-$ 
levels, a property fully reproduced by the DKP model that employs six alternating mass domains 
[Fig.\ \ref{fig1}(d)]. Both our TB and DKP calculations (not shown) confirm that the 
width of the forbidden band decreases exponentially as the length of the hexagonal arm tends to 
infinity, in agreement with earlier findings of a single soliton-antisoliton pair. \cite{jakl83}  

\begin{figure}
\centering\includegraphics[width=6.0cm]{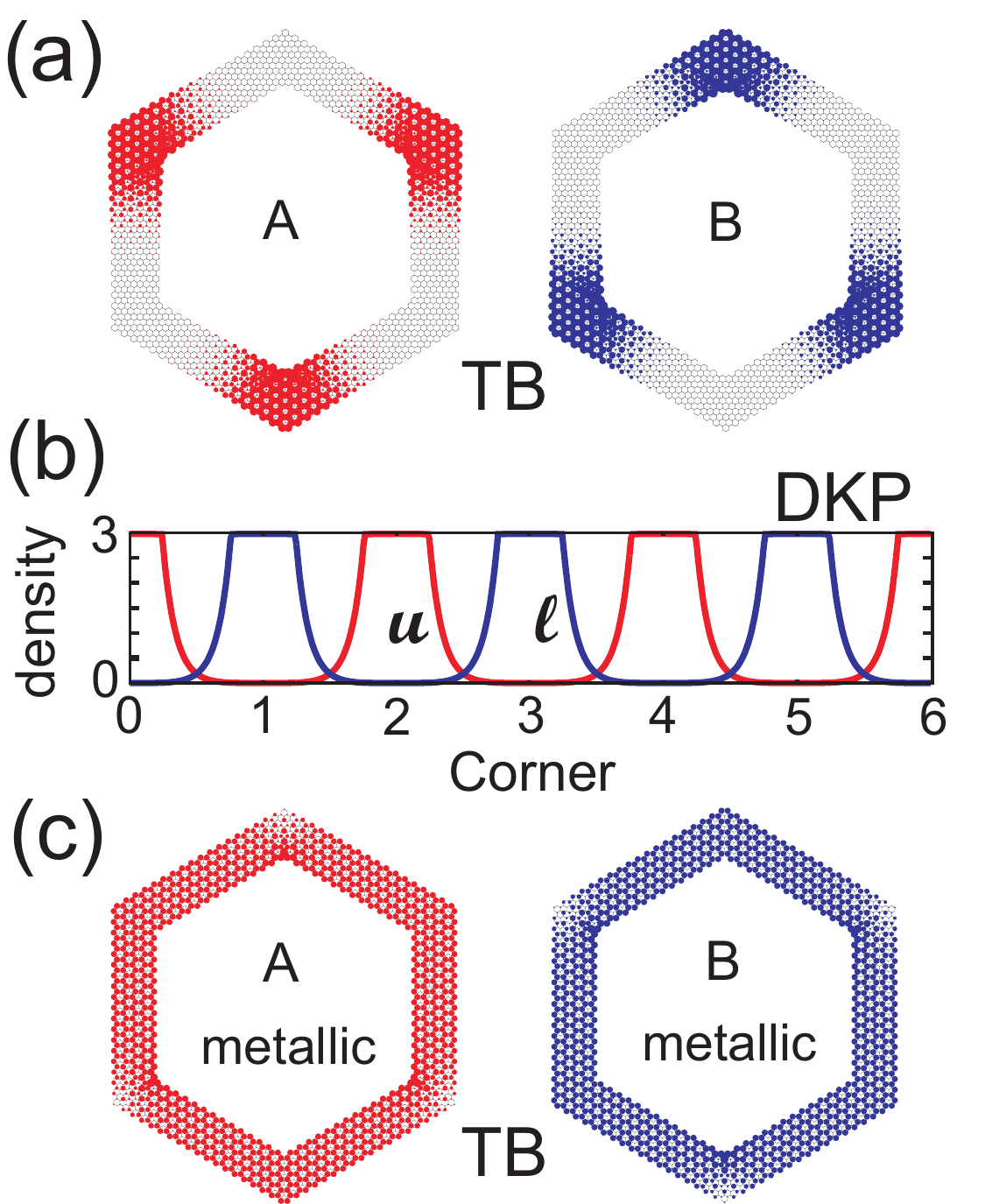}
\caption{
{\it Wave functions for an excitation belonging to the ``forbidden'' solitonic  band.\/} 
(a) $A$-sublattice (red) and $B$-sublattice (blue) components of the TB state with energy
$\varepsilon=0.12507 \times 10^{-2}t$ at $\Phi=\Phi_0/3$, belonging to the forbidden solitonic band
of the class-II nanoring with ${\cal N}_W=16$ [see Fig.\ \ref{fig1}(b)]. 
(b) Upper (red) and lower (blue) spinor components for the corresponding state  
(forbidden band) according to the DKP spectrum [see Fig.\ \ref{fig1}(d)], reproducing 
the TB behavior of the class-II nanoring with ${\cal N}_W=16$ ($m_0=0.18 t/v_F^2$).
The TB and DKP wave functions for all states of the solitonic band are similar to those
displayed here. The wave functions here represent trains of solitons. For contrast, see 
Fig. 10 in Ref.\ \onlinecite{roma13} which portrays schematically the spinor $\Psi_S$ for a 
{\it single\/} fermionic soliton attached to a Higgs field with a smooth kink-soliton 
analytic shape $\phi_k(x)= \phi_0 \tanh \left( \sqrt{\xi/2} \phi_0 x \right)$. 
$\phi_k(x)$ is a solution \cite{jasc81,roma13} of the Lagrangian in Eq.\ (\ref{lagrphi}). 
(c) $A$-sublattice (red) and $B$-sublattice (blue) components of the TB state with energy
$\varepsilon=0.55636 \times 10^{-2}t$ at $\Phi=\Phi_0/3$, associated with the metallic
(class-III) nanoring with ${\cal N}_W=14$ (see corresponding spectrum in Fig.\ 1(b) of
Ref.\ \onlinecite{roma13}). In contrast to the localized-at-the-corners topological-insulator 
wave functions of the semiconducting (Class-II) ring in (a), the metallic-aGRG 
(Class III) wave functions in (c) do not exhibit any localization features  
and are thus devoid of any TI characteristics. DKP densities in units of 
$10^{-3}/h$, where $h=0.35 a_0$.
}   
\label{fig2}
\end{figure}  
  
The strong localization of a fraction of a fermion at the domain walls (hexagon's corners), 
characteristic of fermionic solitons \cite{jasc81} and of soliton/anti-soliton 
pairs, \cite{jakl83} is clearly seen in the TB density distributions (modulus of single-particle 
wave functions) displayed in Fig.\ \ref{fig2}(a). The TB $A$ ($B$) sublattice component localizes 
at the odd (even) numbered corners. These alternating localization patterns (trains of solitons) are 
faithfully reproduced [see Fig.\ \ref{fig2}(b)] by the upper, $\psi_u$, and lower, $\psi_l$, spinor 
components of the DKP model. The three soliton-antisoliton train in Fig.\  
\ref{fig2}(b) generates an unusual $e/6$ charge fractionization at each corner, which is unlike the 
$e/2$ fractionization, familiar from polyacetylene. Moreover, the fractionization patterns in 
topological graphene structures may be tuned. For example, as illustrated below, the more familiar 
$e/2$ fraction \cite{jare76,heeg88,fran09} can be realized in the case of an aGRG with {\it mixed\/} 
class-I and class-III arms.

The absence of a forbidden band (i.e., solitonic excitations within the gap) in the spectrum of the 
class-I hexagonal nanorings [see Fig.\ \ref{fig1}(a), ${\cal N}_W=15$] indicates that the 
corners in this case do not induce an alternation between the two equivalent dimerized domains 
(represented by $\pm m_0$ in the DKP model). Here the corners do not act
as topological domain walls. The inset in Fig.\ \ref{fig1}(c) portrays the DKP 
spectrum when a constant mass ${\cal M}= 0.13 t$ is assumed to encircle the ring. This spectrum 
conforms with that expected from a free massive Dirac fermion, and it clearly disagrees with the TB
spectrum in Fig.\ \ref{fig1}(a). However, direct correspondence between the TB and DKP spectra is 
achieved here too by using a variable Higgs field defined as $\phi(x)=m^{(n)}_i(x)$ with 
$m_1^{(n)}=m_3^{(n)}= 0.06 t/v_F^2$ and
$m_2^{(n)}=0.13 t/v_F^2$ [see the schematic inset in Fig.\ \ref{fig1}(a); the DKP spectrum is plotted
in Fig.\ \ref{fig1}(c)]. $\phi(x)$ exhibits now depressions at 
the hexagon corners, instead of the aforementioned sign alternation; compare insets in Figs.\ 
\ref{fig1}(a) and \ref{fig1}(b). This variation of $\phi(x)$ resembles that of the field used in the 
theory of polarons in conducting polymers, \cite{camp01} and in the theory of fermion bags in hadronic
\cite{camp76} and heavy-quark physics. \cite{mapa90} 

\begin{figure}
\centering\includegraphics[width=7.0cm]{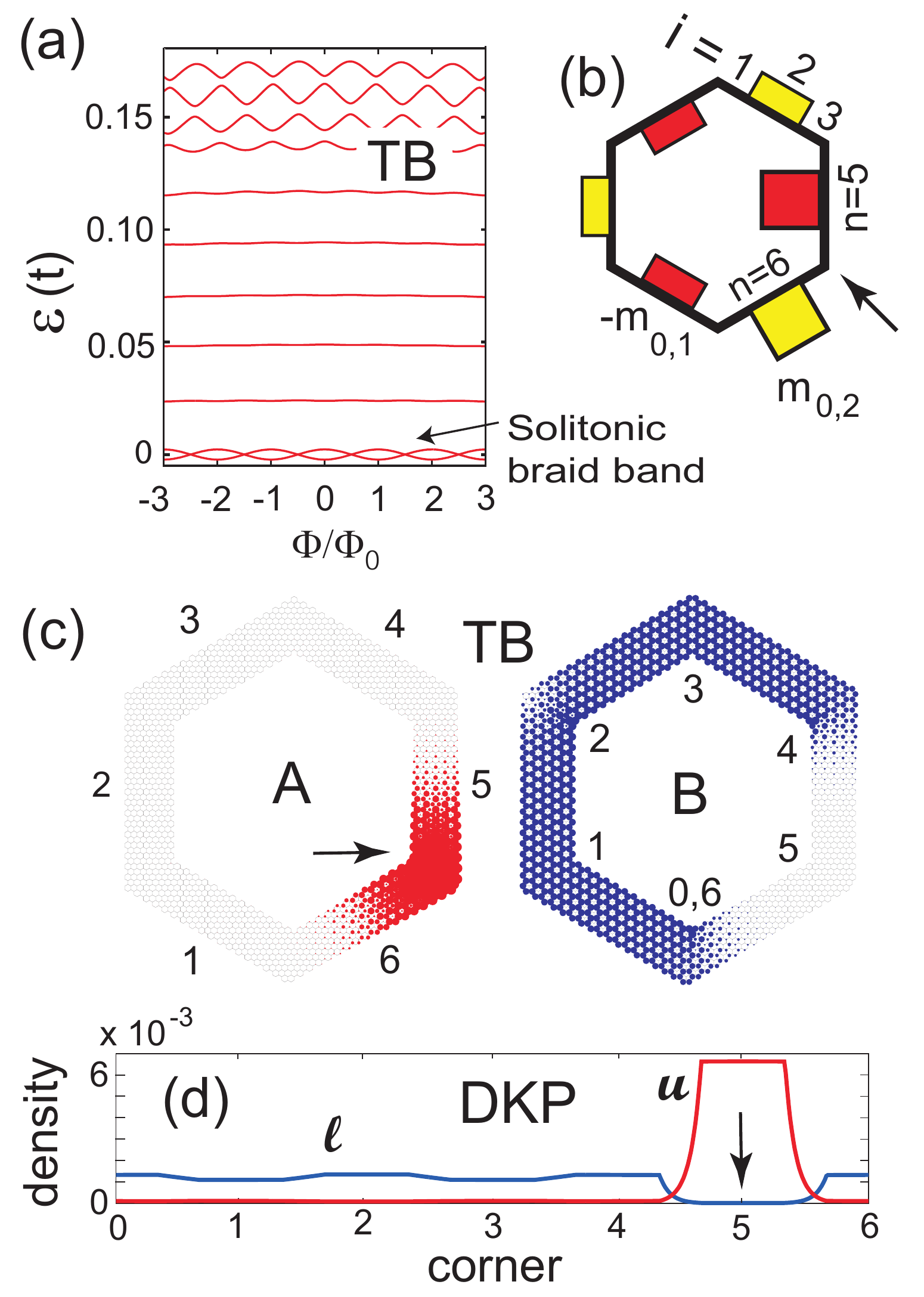}
\caption{
{\it Spectra and wave functions of a mixed ring with arms belonging to two different 
classes.\/} Two contiguous arms [5 and 6 in (b)] belong to class I (semiconducting ribbons, 
${\cal N}_W=15$) and the remaining arms belong to class III (metallic ribbons, ${\cal N}_W=17$).
(a) The TB  Aharonov-Bohm spectrum. 
(b) Schematic of the $\phi(x)$ field in the DKP model, yielding the best reproduction [not shown, but 
see the wave function in (d)] of the TB spectra in (a). Note the unequal size of the colored boxes
specifying $\phi(x)$. The $m_{0,2}$ mass (class-I arms), is much larger than the $m_{0,1}$ one
(class-III arms). Positive (negative) mass values are indicated in yellow (red). 
(c) TB wave function for a state of the braid band at $\Phi=\Phi_0/3$ 
($\varepsilon=0.1995 \times 10^{-2} t$), for the $A$ (red) and $B$ (blue) sublattice components.
(d) The upper (red) and lower (blue) spinor components of the DKP state, corresponding 
and showing agreement with the TB state in (c).
The arrows [in (b), (c), and (d)] indicate the single hexagon corner where the soliton is 
localized. Parameters for the DKP model are: 
$m_1^{(n)}=m_3^{(n)}=0$, and $m_2^{(n)}=(-1)^n m_{0,1}$, for $n=1,2,3,4$ and 
$m_2^{(n)}=(-1)^n m_{0,2}$ for $n=5,6$, with $m_{0,1}=0.01 t/v_F^2$ and $m_{0,2}=0.20 t/v_F^2$.
$a=10 a_0$ and $b=10 a_0$. In (c) the indices denote the arms (left) and corners (right). 
DKP densities in units of $1/h$, where $h=0.36 a_0$.
}
\label{fig3}
\end{figure}  

\subsection{Rings with mixed semiconducting/metallic arms}

The pure metallic-aGRG (Class III) wave functions do not exhibit any localization features and are 
thus devoid of any topological-insulator characteristics; see Fig.\ \ref{fig2}(c). Unique TI 
configurations, however, can be formed in mixed rings, i.e., with arms belonging to
different classes. Fig.\ \ref{fig3} portrays a mixed ring, with four arms 
belonging to class-III (${\cal N}_W=17$; metallic) and the two remaining ones to class-I 
(${\cal N}_W=15$; semiconducting) ribbons. The TB spectra are displayed in Fig.\ \ref{fig3}(a) and 
Fig.\ \ref{fig3}(b) describes schematically the Higgs field $\phi(x)=m_i^{(n)}(x)$, which yields the 
best DKP reproduction of the TB spectra (see caption). 
The TB spectrum in Fig.\ \ref{fig3}(a) reflects the loss of sixfold symmetry of the Higgs field
(in contrast to Fig.\ \ref{fig1}). Furthermore, five states in the energy range 
$0.01 t/v_F^2 < \varepsilon < 0.13 t/v_F^2$ exhibit a magnetic-field independent 
flat profile, corresponding to the behavior of a particle-in-a-box. 
Namely, the practically ($m_{0,1}=0.01t/v_F^2$) massless Dirac fermion is 
confined in the potential well formed by the four arms $n=1$ to 4, unable to penetrate under the high
barrier represented by the larger masses $\pm m_{0,2}= \pm 0.20 t/v_F^2$ associated with the fifth and
six arms of the hexagon. The two-fold braid band around $\varepsilon=0$
exhibits a clear Aharonov-Bohm dependence on the magnetic flux $\Phi$. The TB wave function of
one state in this band (with $\varepsilon=0.1995 \times 10^{-2} t$ at $\Phi=\Phi_0/3$) is plotted in 
Fig.\ \ref{fig3}(c). It describes the emergence of a fermionic soliton (with $e/2$ fractional charge)
localized at the domain wall (corner denoted by an arrow) between the fourth and the fifth arms of 
the hexagon. The DKP modeling closely reproduces this TB solitonic wave function, as seen from the 
densities of the upper (red) and lower (blue) spinor components of the fermionic field $\Psi$. 

A central finding of the paper concerns the emergence of topological insulator \cite{haka10,zhan11} 
aspects in certain classes of semiconducting, as well as of mixed metallic-semiconducting, armchair 
graphene nanorings. Indeed it is well established that the Su-Schrieffer-Heeger 
(SSH) model \cite{heeg88} for polyacetylene (and its Jackiw-Rebbi RQF counterpart \cite{jare76}) is
\cite{gura10,ryu11,shenbook,atal12} a two-band nontrivial one-dimensional TI.  In particular, the 
topological domain with a positive mass $m_0>0$ is a trivial insulator with a Chern number equal to 
zero, while the topological domain with $m_0<0$ is a nontrivial TI with a Chern number equal to unity. 
The localized fermionic kink solitons [Figs.\ \ref{fig2}(a), \ref{fig2}(b), and \ref{fig3}] at the 
domain walls (corners of the hexagonal aGRGs connecting adjacent arms, i.e., domains
with different Chern numbers) correspond to the celebrated TI edge states (end states\cite{shenbook} 
for 1D systems), used as a fingerprint for the emergence of the TI state. Usually, realization of a 
TI requires consideration of the spin-orbit coupling, which however is negligible for planar graphene. 
Currently, attempts to enhance the spin-orbit coupling of graphene via adatom deposition is attracting 
attention. \cite{week11} The present findings point to a different direction for realizing a 
graphene-based TI through the manipulation of the geometry of the honeycomb lattice, which is able to 
ovecome the drawback of negligible spin-orbit coupling.

\section{Summary}

In summary, we have advanced and illustrated that the doubly-connected, polygonal geometry of 
graphene rings brings forth, in addition to the celebrated Aharonov-Bohm physics, 
\cite{imry83,roma12} an as-yet unexplored platform spawning topological arrangements 
(including in particular realization of 1D nontrivial topological insulators) for accessing 
acclaimed one-dimensional relativistic quantum field models. \cite{jare76,jasc81,camp76,mapa90} These 
include generation of position-dependent masses, solitonic excitations, and charge fractionization, 
beyond the constant-mass Dirac and DW fermions. These intriguing phenomena, coupled with advances in
preparation of atomically precise graphene nanostructures, \cite{cai10,note2} artificial forms of 
graphene, \cite{pell09,mano12} topological insulators, \cite{haka10,zhan11} and graphene mimics in 
ultracold-atom optical lattices, \cite{tarr12} provide impetus \cite{fuhr13,note3} for further 
experimental and theoretical endeavors.

\begin{acknowledgments}
This work is supported by the Office of Basic Energy Sciences of the US D.O.E. (FG05-86ER45234).
\end{acknowledgments}

\appendix

\section{Tight-binding method}
\label{tbm}

To calculate the single-particle spectrum [the energy levels $\varepsilon_i(\Phi)$] of the graphene
nanorings in the tight-binding approximation, we use the hamiltonian
\begin{equation}
H_{\text{TB}}= - \sum_{<i,j>} \tilde{t}_{ij} c^\dagger_i c_j + h.c.,
\label{htb}
\end{equation}
with $< >$ indicating summation over the nearest-neighbor sites $i,j$. The hopping parameter
\begin{equation}
\tilde{t}_{ij}=t \exp \left( \frac{ie}{\hbar c}  \int_{{\bf r}_i}^{{\bf r}_j}
d{\bf s} \cdot {\bf A} ({\bf r}) \right),
\label{tpei}
\end{equation}
where ${\bf r}_i$ and ${\bf r}_j$ are the positions of the carbon atoms $i$ and $j$, respectively,
and ${\bf A}$ is the vector potential (in the Landau gauge) associated with the constant magnetic
field $B$ applied perpendicularly to the plane of the nanoring. $\Phi= BS$ is the magnetic flux
through the area $S$ of the graphene ring and $\Phi_0 = hc/e$ is the flux quantum. $t=2.7$ eV
is the hopping parameter of the two-dimensional graphene.

The derivation of the effective 1D tight-binding equation for an aGR, given in Ref.\
\onlinecite{zhen07} [see Eq.\ (6) therein], starts with the 2D TB Hamiltonian here [Eq.\ (\ref{htb})
above] and involves Fourier expansions of the wave functions of the $A$ and $B$ sublattices.

\section{Dirac-Kronig-Penney superlatice model}
\label{dkpsm}

The building block of the DKP model is a 2$\times$2 wave-function matrix ${\bf \Omega}$ formed by the
components of two independent spinor solutions (at a point $x$) of the onedimensional, first-order
generalized Dirac equation [see Eq.\ (\ref{direq}) above]. 
${\bf \Omega}$ plays \cite{mcke87} the role of the Wronskian matrix
${\bf W}$ used in the second-order nonrelativistic Kronig-Penney model. Following Ref.\
\onlinecite{mcke87}, we use the simple form of ${\bf \Omega}$ in the Dirac representation
($\alpha=\sigma_1$, $\beta=\sigma_3$), namely
\begin{equation}
{\bf \Omega}_K (x) = \left( \begin{array}{cc}
e^{i K x} & e^{-i K x} \\
\Lambda e^{i K x}& -\Lambda e^{-i K x} \end{array} \right),
\label{ome}
\end{equation}
where
\begin{equation}
K^2=\frac{(E-V)^2-m^2 v_F^4}{\hbar^2 v_F^2}, \;\;\; \Lambda= \frac{\hbar v_F K}{E-V+m v_F^2}.
\label{klam}
\end{equation}
The transfer matrix for a given region (extending between two matching points $x_1$ and $x_2$
specifying the potential steps $m_i^{(n)}$) is the product
${\bf M}_K (x_1,x_2)= {\bf \Omega}_K (x_2){\bf \Omega}_K^{-1} (x_1)$;  this latter matrix depends
only on the width $x_2-x_1$ of the region, and not separately on $x_1$ or $x_2$.

The transfer matrix corresponding to the $n$th arm of the hexagon can be formed \cite{roma13} 
as the product
\begin{equation}
{\bf t}_n = \prod_{i=1,3} {\bf M}_K (x_i,x_{i+1}),\;\;\; x_1=0,\; x_4=L,
\label{tside}
\end{equation}
with $L$ being the (common) length on the hexagon arm. The transfer matrix associated with the
complete unit cell (encircling the hexagonal ring) is the product
\begin{equation}
{\bf T}=\prod_{n=1}^6 {\bf t}_n.
\label{thex}
\end{equation}

Following  Refs.\ \onlinecite{roma13,imry83}, we consider the superlattice generated from the virtual
periodic translation of the unit cell as a result of the application of a magnetic field $B$
perpendicular to the ring. Then the Aharonov-Bohm energy spectra are given as solutions of the
dispersion relation
\begin{equation}
\cos \left[ 2\pi(\Phi/\Phi_0+1/2) \right] = \Tr[{\bf T}(E)]/2,
\label{disrel}
\end{equation}
where we have explicitly denoted the dependence of the r.h.s. on the energy $E$.

The energy spectra and single-particle densities do not depend on a specific representation.
However, the wave functions (upper and lower spinor components of the fermionic field $\Psi$) do
depend on the representation used. To transform the initial DKP wave functions to the
($\alpha=\sigma_2$, $\beta=\sigma_1$) representation, which corresponds to the natural separation of
the tight-binding amplitudes into the $A$ and $B$ sublattices, we apply successively the unitary
transformations $D_{23}=(\sigma_2+\sigma_3)/\sqrt{2}$ and $D_3=\exp(i \pi \sigma_3/4)$.


\newpage
\end{document}